\documentclass[aps,prc,amsmath,preprint,superscriptaddress,showkeys,showpacs]{revtex4-1}
\usepackage{graphicx}
\usepackage{color}
\usepackage{epsfig,epstopdf}
\newcommand{\beq}{\begin{equation}}
\newcommand{\eeq}{\end{equation}}
\newcommand{\beqy}{\begin{eqnarray}}
\newcommand{\eeqy}{\end{eqnarray}}

\begin{document}

\title{Role of the symmetry energy on the neutron-drip transition in accreting and nonaccreting neutron stars}
\author{A.~F. Fantina}
\affiliation{Institut d'Astronomie et d'Astrophysique, CP-226, Universit\'e Libre de Bruxelles (ULB),
1050 Brussels, Belgium}
\author{N. Chamel}
\affiliation{Institut d'Astronomie et d'Astrophysique, CP-226, Universit\'e Libre de Bruxelles (ULB),
1050 Brussels, Belgium}
\author{Y. D. Mutafchieva}
\affiliation{Institute for Nuclear Research and Nuclear Energy, Bulgarian Academy of Sciences, 72 Tsarigradsko Chaussee, 1784 Sofia, Bulgaria}
\author{Zh. K. Stoyanov}
\affiliation{Institute for Nuclear Research and Nuclear Energy, Bulgarian Academy of Sciences, 72 Tsarigradsko Chaussee, 1784 Sofia, Bulgaria}
\author{L. M. Mihailov}
\affiliation{Institute of Solid State Physics, Bulgarian Academy of Sciences, 72 Tsarigradsko Chaussee, 1784 Sofia, Bulgaria}
\author{R. L. Pavlov}
\affiliation{Institute for Nuclear Research and Nuclear Energy, Bulgarian Academy of Sciences, 72 Tsarigradsko Chaussee, 1784 Sofia, Bulgaria}

\date{\today}

\begin{abstract}
In this paper, we study the role of the symmetry energy on the neutron-drip transition in both nonaccreting and accreting neutron stars, allowing 
for the presence of a strong magnetic field as in magnetars. 
The density, pressure, and composition at the neutron-drip threshold 
are determined using the recent set of the Brussels-Montreal microscopic nuclear mass models, which mainly differ in their predictions for the value of the symmetry energy 
$J$ and its slope $L$ in infinite homogeneous nuclear matter at saturation. Although some correlations between on the one hand the neutron-drip density, the pressure, the proton fraction and 
on the other hand $J$ (or equivalently $L$) are found, these correlations are radically different in nonaccreting and accreting neutron stars. In particular, 
the neutron-drip density is found to increase with $L$ in the former case, but decreases in the latter case depending on the composition of ashes from 
x-ray bursts and superbursts. We have qualitatively explained these different behaviors using a simple mass formula. We have also shown that the details 
of the nuclear structure may play a more important role than the symmetry energy in accreting neutron-star crusts. 
\end{abstract}


% symmetry energy, neutron stars, NS crust, EoS of NS matter

\maketitle

\section{Introduction}
\label{sec:intro}

Born from the gravitational core collapse of massive stars in supernova explosions, neutron stars are the densest known stars in the universe (see, e.g., 
Ref.~\cite{haensel2007}). Most of them have been discovered as isolated radio pulsars. Neutron stars can also belong to binary systems, in which the 
neutron star accretes matter from a stellar companion. These systems are observed as x-ray pulsars. Neutron stars may be endowed with very strong magnetic 
fields. Soft $\gamma$-ray repeaters 
and anomalous x-ray pulsars are thought to be such so-called ``magnetars'' (see, e.g., Ref.~\cite{woods2006} for a review). Their surface magnetic field, as inferred 
from spin-down and spectroscopic studies, are of the order of $10^{14}-10^{15} $~G \cite{olausenkaspi2014, tiengo2013, hongjun2014}. Numerical simulations 
suggest that their internal magnetic field could be even stronger, reaching $10^{18}$~G~\cite{pili2014,chatterjee2015}. 

Apart from a thin atmospheric plasma layer of light elements (mainly hydrogen and helium) possibly surrounding a Coulomb liquid of electrons and ions, a 
neutron star is thought to contain at least three distinct regions. The outermost region consists of a solid crust made of a crystal lattice of fully ionized 
atoms arranged on a body-centered cubic lattice (see, e.g., Ref.~\cite{chamelhaensel2008} for a review). With increasing density, nuclei become progressively 
more neutron rich due to electron captures until neutrons start to drip out of nuclei at some threshold density $\rho_{\rm drip}$, thus marking the boundary 
between the outer crust and the inner crust. In unmagnetized or weakly magnetized nonaccreting neutron stars (with magnetic fields $\ll 10^{14}$~G), the neutron-drip 
transition is found to occur at density $\rho_{\rm drip} \approx 4 \times 10^{11}$~g~cm$^{-3}$ (see, e.g., Refs.~\cite{bps1971, haensel2007, roca2008, pearson2011, 
wolf2013, kreim2013, chamel2015c}). On the other hand, this density can be shifted in magnetars, due to the presence of strong magnetic fields (see, e.g., 
Refs.~\cite{sek1977, lai91, nandi2011, chamel2012, nandi2013, vishal2014, chamel2015b,bas2015}), and in accreting neutron stars, due to the accretion from a companion 
\cite{hz1990, hz2003, chamel2015a}.
The crust dissolves into an homogeneous liquid mixture at about half the density prevailing in heavy atomic nuclei. This transition is expected to be realized 
through different nuclear structures, such as cylinders, rods, and plates (also called ``nuclear pasta'', see Ref.~\cite{haensel2007,chamelhaensel2008} and references therein). Further 
inside, at very high densities in the neutron-star core, additional degrees of freedom, such as hyperons, meson condensates, and/or deconfined quarks, could be 
present (see, e.g., Ref.~\cite{haensel2007, page2006, weber2007}).

From the nuclear physics point of view, the neutron-star crust is also a unique ``laboratory'' to probe the properties of infinite homogeneous asymmetric nuclear matter, in particular 
the symmetry energy, at subsaturation densities (see, e.g. Ref.~\cite{steiner2015}). The symmetry energy is generally defined as 
\beq
\label{eq:e-inm}
S_1(n) = \frac{1}{2}\frac{\partial^2 (\mathcal{E}/n)}{\partial \eta^2}\biggr\vert_{\eta=0} \ ,
\eeq
where $\mathcal{E}(n,\eta)$ is the energy density of infinite homogeneous nuclear matter with proton (neutron) density $n_p$ ($n_n$), baryon density 
$n=n_n + n_p$, and charge asymmetry $\eta = (n_n-n_p)/n$. Alternatively, the symmetry energy often refers to the difference between the energy of 
pure neutron matter and that of symmetric matter:
\beq
\label{eq:s2}
S_2(n) =\frac{\mathcal{E}(n,\eta=1) - \mathcal{E}(n,\eta=0)}{n} \ .
\eeq 
Because $\mathcal{E}(n,\eta)$ generally contains terms of order $\eta^4$ and higher, the two definitions $S_1(n)$ and $S_2(n)$ of the symmetry energy 
do not exactly coincide (see, e.g.  the discussion in Sec.~III of Ref.~\cite{goriely2010}). From now on, we adopt the first definition. 
The symmetry energy of infinite homogeneous nuclear matter $S_1(n)$ can be expanded around the saturation density $n_0$ (whose value is $\approx 0.16$~fm$^{-3}$) as
\beq
\label{eq:s1}
S_1(n) \approx J + \frac{1}{3} L \left( \frac{n-n_0}{n_0} \right) + \frac{1}{18} K_{\rm sym} \left( \frac{n-n_0}{n_0} \right)^2\ .
\eeq
Whereas the value of the symmetry energy at saturation, $J$, is fairly well constrained by nuclear physics experiments to lie around 30~MeV, the values of the 
slope of the symmetry energy, $L$, and of higher order coefficients like $K_{\rm sym}$ at saturation,
 are still very uncertain and poorly constrained (see, e.g., the discussion in 
Refs.~\cite{tsang2012,lattimerlim2013}). The symmetry energy has been shown to affect the composition of neutron-star crusts and the crust-core transition~(see, e.g., 
Refs.~\cite{horowitz2001,oyamatsu2007, roca2008, vidana2009,ducoin2011,grill2012,newton2013,sulaksono2014,provid2014,bao2014, grill2014,seif2014,provid2014}). 
Comparatively few studies have been devoted to the role of the symmetry energy on the boundary between the outer and inner crusts~\cite{bao2014} (some discussions can 
also be found, e.g., in Refs.~\cite{roca2008,provid2014,grill2014}). Moreover, all these studies focused on nonaccreting and unmagnetized neutron stars.

In this paper, we study the role of the symmetry energy on the onset of neutron drip in both nonaccreting and accreting neutron stars, allowing for the 
presence of a strong magnetic field as in magnetars. To this end, we have made use of the recent set of Brussels-Montreal microscopic nuclear mass 
models~\cite{goriely2013}. Our model of neutron-star crust is presented in Sec.~\ref{sec:model}~: after discussing our assumptions in Sec.~\ref{sec:assumptions}, 
the nuclear mass models are briefly described in Sec.~\ref{sec:hfb-models}. The determination of the neutron-drip transition is discussed in Sec.~\ref{sec:neutron-drip}, 
and numerical results are presented in Sec.~\ref{sec:results}.

% MODEL OF NS CRUST
\section{Model of neutron-star crust}
\label{sec:model}

\subsection{Main assumptions}
\label{sec:assumptions}

In the region of the crust that we consider here, the pressure is high enough that atoms are fully ionized~\cite{chamelhaensel2008}. We assume that the temperature $T$ is 
lower than the crystallization temperature $T_m$ so that nuclei are arranged in a regular crystal lattice. Considering that the crystalline structures are 
made of only one type of ion $^{A} _{Z} X$, with mass number $A$ and atomic number $Z$, the crystallization temperature $T_m$ is given by 
(see, e.g., Ref.~\cite{haensel2007}):
\beq
T_m = \frac{e^2}{a_e k_{\rm B} \Gamma_m} Z^{5/3} \ ,
\eeq
where $e$ is the elementary electric charge, $a_e=(3/(4\pi n_e))^{1/3}$ the electron-sphere radius expressed in terms of the electron number density $n_e$, 
$k_\text{B}$ is the Boltzmann's constant, and $\Gamma_m\simeq 175$ is the Coulomb coupling parameter at melting. Typically, $T_m$ is much lower than the 
electron Fermi temperature defined by
\beq
 T_{\rm F}=\frac{\mu_e-m_e c^2}{k_{\rm B}} \ ,
\eeq
where $m_e$ is the electron mass, $c$ is the speed of light, and $\mu_e$ is the electron Fermi energy. 
Therefore, electrons are highly degenerate and from now on we will set $T=0$. To a very good approximation, electrons can be treated as an ideal relativistic 
Fermi gas. Expressions for the electron energy density $\mathcal{E}_e$ and electron pressure $P_e$ can be found in Ref.~\cite{haensel2007}. 
The main corrections arise from electron-ion interactions (see, e.g., Ref.~\cite{pearson2011} and references therein for a discussion of other corrections). 
Neglecting the finite size of ions and the quantum zero-point motion of ions off their equilibrium position, the lattice contribution to the energy density 
is given by 
(see e.g. Chap. 2 in Ref.~\cite{haensel2007})
\beq
\label{eq:EL}
\mathcal{E}_L  = C e^2 n_e^{4/3} Z^{2/3} \ ,
\eeq
where $C$ is a crystal structure constant, whose value for a body-centered cubic lattice is $C=-1.444231$~\cite{baiko2001}. This expression still remains valid in the 
presence of a strong magnetic field, as a consequence of the Bohr-van Leeuwen theorem~\cite{vanvleck1932}. 
The matter pressure $P$ can be expressed as %the sum of the electron and lattice contribution (indeed, nuclei exert no pressure at zero temperature):
\beq
\label{eq:ptot}
P = P_e + P_L \ ,
\eeq
where $P_e$ is the pressure of a uniform electron gas, and the lattice pressure $P_L$ is given by 
\beq
\label{eq:PL}
 P_L=\frac{\mathcal{E}_L}{3} \ .
\eeq
The only microscopic inputs for the description of the outer crust are nuclear masses $M^\prime(A,Z)$. They can be obtained from the corresponding 
atomic mass $M(A,Z)$ after subtracting out the binding energy of the atomic electrons (see Eq.~(A4) of Ref.~\cite{lpt03}):
\begin{equation}
\label{3}
M^\prime(A,Z)c^2 = M(A,Z)c^2 + 1.44381\times 10^{-5}\,Z^{2.39} + 1.55468\times 10^{-12}\,Z^{5.35} \ ,
\end{equation}
where both masses are expressed in units of MeV$/c^2$. Nuclear masses may be changed in the presence of a strong magnetic field. On the other hand, 
for the magnetic field strength we shall consider, namely $B < 10^{17} $~G, those changes are very small \cite{pena2011} and will thus be ignored. 
For the masses that have not yet been measured, we have made use of the microscopic mass tables computed by the Brussels-Montreal group  (for a recent review of these models, see, e.g., Ref.~\cite{chamel2015c}). 

\subsection{Microscopic nuclear mass models}
\label{sec:hfb-models}

The family of Brussels-Montreal nuclear mass models that we adopt here~\cite{goriely2013} are based on the nuclear energy density functional (EDF) 
theory using a generalized form of Skyrme zero-range effective interactions~\cite{chamel2009}, supplemented with a microscopic contact pairing interaction~\cite{chamel2010}. 
For all these models, the masses of nuclei were obtained by adding to the 
Hartree-Fock-Bogoliubov (HFB) energy a phenomenological Wigner term and a correction term for the rotational and vibrational spurious collective energy 
(see, e.g. Ref.~\cite{goriely2010} for a discussion about the accuracy of this latter correction). The EDFs BSk22, BSk23, BSk24, BSk25, and BSk26 
underlying the nuclear mass models HFB-22, HFB-23, HFB-24, HFB-25, and HFB-26 were primarily fitted to the 2353 measured masses of nuclei with $N$ and $Z \geq 8$ from 
the 2012 Atomic Mass Evaluation \cite{audi2012}, with a root-mean-square (rms) deviation of $0.63$~MeV, $0.57$~MeV, $0.55$~MeV, $0.54$~MeV, and $0.56$~MeV 
respectively. At the same time, these EDFs were constrained to reproduce the equation of state (EoS) of homogeneous neutron matter, as obtained by many-body 
calculations using realistic two- and three-nucleon interactions. Moreover, the incompressibility $K_v$ of infinite homogeneous symmetric nuclear matter at saturation was 
required to fall in the range $240\pm10$~MeV~\cite{colo2004}, and the isoscalar effective mass was fixed to the realistic value $M_s^*=0.8$. In addition, the 
EoS of symmetric nuclear matter obtained from these EDFs were found to be compatible with the empirical constraints inferred from the analysis of heavy-ion 
collision experiments~\cite{dan02,lynch09}. For all these reasons, we believe that these EDFs can be reliably applied to the description of neutron-star crusts. 

\begin{figure*}
\begin{center}
\includegraphics[scale=0.35]{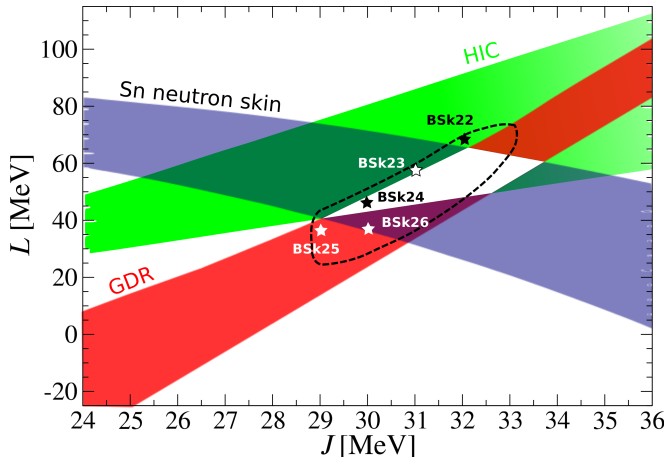}
\end{center}
%\vskip -0.5cm
\caption{(Color online) Experimental constraints on the symmetry energy parameters, taken from Ref.\cite{lattimerlim2013}; the dashed line represents the constraint 
obtained from fitting experimental nuclear masses using the Brussels-Montreal Hartree-Fock-Bogoliubov models (including unpublished ones) with an root-mean-square deviation below 
$0.84$~MeV; star symbols correspond to the series of models from Ref.~\cite{goriely2013} that we consider in this work. See text for details.}
\label{fig:JvsL}
\end{figure*}

In generating these five EDFs, different values of the symmetry energy coefficient $J$ 
were imposed, thus making them suitable for a systematic study of 
the role of the symmetry energy on the outer crust of a neutron star. In particular, BSk22, BSk23, BSk24 and BSk25 were constrained to the symmetry energy 
coefficients $J = 32$, $31$, $30$ and $29$~MeV, respectively and were all fitted to the realistic neutron-matter EoS labeled ``V18" in Ref.~\cite{ls2008}, 
while BSk26 was fitted to the EoS labeled ``A18 + $\delta\,v$ + UIX$^*$"in Ref.~\cite{apr1998} under the constraint $J = 30$~MeV. The values of the symmetry energy coefficient $J$, as well as the higher-order coefficients 
$L$ and $K_{\rm sym}$ for the different EDFs are indicated in Table \ref{tab:esymcoef}. As already discussed in Ref.~\cite{pearson2014}, $L$ is strongly 
correlated with $J$: increasing $J$ leads to higher values of $L$. The difference between the values of $L$ obtained with BSk24 and BSk26 (for which $J=30$ MeV) 
arises from the constraining neutron-matter EoS. Indeed, as shown in Eqs.~(\ref{eq:e-inm}) and (\ref{eq:s1}) the softer the underlying neutron-matter EoS, 
the lower $L$. On the other hand, the correlation between $L$ and $K_{\rm sym}$ is less clear. As shown in Fig.~\ref{fig:JvsL}, the values of $J$ and $L$ obtained 
with the Brussels-Montreal EDFs are consistent with constraints coming from the combined analysis of various experiments \cite{tsang2012, lattimerlim2013} 
(see also the discussion in Sec.~IIIC in Ref.~\cite{goriely2013}): the constraint deduced in Ref.~\cite{tsang2009} from heavy-ion collisions (HIC), the constraint derived 
in Ref.~\cite{chen2010} from measurements of the neutron-skin thickness in tin isotopes, and finally the constraint obtained from the analysis of the giant dipole 
resonance (GDR) \cite{trippa2008}. 

\begin{figure*}
\begin{center}
\includegraphics[scale=0.35]{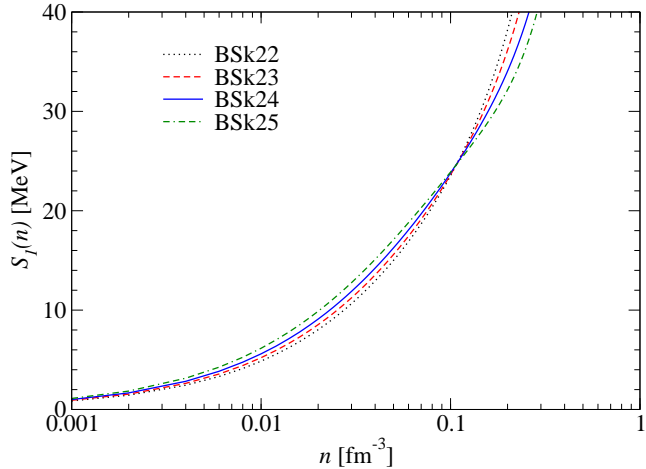}
\end{center}
%\vskip -0.5cm
\caption{(Color online) Symmetry energy of infinite homogeneous nuclear matter $S_1(n)$ versus density, 
for the Brussels-Montreal energy density functionals BSk22 to BSk25~\cite{goriely2013}.}
\label{fig:esym-vs-n}
\end{figure*}

The density dependence of the symmetry energy $S_1(n)$, Eq.~(\ref{eq:s1}), for the Brussels-Montreal EDFs BSk22 to BSk25 is shown in Fig.~\ref{fig:esym-vs-n}. The most notable feature, due to the mass fit, 
is the crossing of all curves at densities around $0.11$~fm$^{-3}$. Indeed, the binding energy of finite nuclei is mainly sensitive to the symmetry energy 
at such densities rather than to the symmetry energy at saturation density~\cite{horowitz2001, zhangchen2013}. 
For these functionals, which were fitted to the same 
neutron-matter EoS, we observe a clear correlation between the density dependence of the symmetry energy and its value at saturation with a change of hierarchy 
at the crossing point: the higher $J$, the higher (lower) the symmetry energy above (below) this point. From these considerations, the ``effective'' value of 
the symmetry energy averaged over the volume of a nucleus is thus expected to increase with decreasing $J$, as found in previous studies~\cite{goriely2005,gaidarov2012}.

Comparing various constraints from both nuclear physics and astrophysics~\cite{pearson2014, fantina2015}, BSk24 (BSk22) was found to be the best (worst) in the series of Brussels-Montreal EDFs BSk22-BSk26. 
In the following, we shall thus take BSk24 as the reference EDF. We will not consider BSk26 since it was fitted to a different neutron-matter EoS from the other EDFs. 
We have made use of the mass tables from the BRUSLIB database~\cite{bruslib}.

\begin{table}
\centering
\caption{Symmetry energy coefficient $J$ and higher-order symmetry energy coefficients of infinite 
homogeneous nuclear matter at saturation for the Brussels-Montreal energy density functionals~\cite{goriely2013}.}\smallskip
\label{tab:esymcoef}
\begin{tabular}{|c|ccccc|}
\hline %\noalign {\smallskip}
 & BSk22 & BSk23 & BSk24 & BSk25 & BSk26 \\
\hline %\noalign {\smallskip}
%$n_0$ [fm$^{-3}$] & 0.1578 & 0.1578 & 0.1578 & 0.1587 & 0.1589 \\
$J$ [MeV] & 32.0 & 31.0 & 30.0 & 29.0 & 30.0 \\
$L$ [MeV] & 68.5 & 57.8 & 46.4 & 36.9 & 37.5  \\
$K_{\rm sym}$ [MeV] & 13.0 & -11.3 & -37.6 & -28.5 & -135.6 \\
%$a_v$ [MeV] & -16.088 & -16.068 & -16.048 & -16.032 & -16.064 \\
%$K_v$ & 245.9 & 245.7 & 245.5 & 236.0 & 240.8 \\
\hline
\end{tabular}
\end{table}

\section{Neutron-drip transition}
\label{sec:neutron-drip}

Neutron stars are formed from the catastrophic gravitational core collapse of massive stars during supernova explosions. Under such extreme conditions, it is generally assumed that 
all kinds of nuclear and electroweak processes occur and that, as the neutron star cools down, matter remains in full thermodynamic equilibrium. Eventually, 
the neutron star becomes cold and fully ``catalyzed'' \cite{haensel2007} unless matter is accreted from a companion star. In the latter case, the accretion 
may heat the neutron star and change the composition of its crust. These two different astrophysical scenarios will thus be treated separately.

\subsection{Nonaccreting neutron stars}

The usual procedure~\cite{bps1971} to determine the equilibrium nucleus in any layer of the outer crust of a nonaccreting neutron star at pressure $P$ is to minimize the Gibbs 
free energy per nucleon defined by 
\beq
\label{eq:gibbs-general}
g = \frac{\mathcal{E}+P}{n} \ ,
\eeq
where $\mathcal{E}$ denotes the mean energy density of the crustal matter, and $n$ denotes the mean baryon number density. As shown in Ref.~\cite{chamel2015d}, $g$ 
remains the suitable thermodynamic potential in the presence of a strong magnetic field. Equation~(\ref{eq:gibbs-general}) can be equivalently expressed 
as~\cite{bps1971,lai91} 
\begin{equation}
\label{eq:gibbs}
g=\frac{M'(A,Z)c^2}{A}+\frac{Z}{A}\left(\mu_e -m_e c^2+\frac{4}{3} C e^2 n_e^{1/3} Z^{2/3}\right)\, .
\end{equation}
Ignoring neutron band structure effects~\cite{chamel2006,chamel2007}, 
the onset of neutron drip is determined by the condition $g=m_n c^2$, where $m_n$ is the neutron mass~\cite{bps1971} (see also the discussion in Ref.~\cite{chamel2015a}). 
This latter condition can be approximately expressed as \cite{chamel2015a}
\beq
\label{eq:n-drip-mue}
\mu_e(n_e^{\rm drip}) + \frac{4}{3}C e^2 (n_e^{{\rm drip}})^{1/3} Z^{2/3} =  \mu_e^{{\rm drip}}\, ,
\eeq
where
\beq
\label{eq:muedrip}
\mu_e^{{\rm drip}}(A,Z)\equiv \frac{-M'(A,Z)c^2+A m_n c^2}{Z} +m_e c^2 \, .
\eeq
Note that this condition remains the same in the presence or in the absence of a magnetic field. 
However, since the relation between the electron density and 
electron Fermi energy does depend on the magnetic field, the neutron-drip density and pressure do depend on the magnetic field. 

For weakly magnetized neutron stars, $B\lesssim B_\star$ where $B_\star \equiv B/B_{\rm crit}$ is the magnetic field expressed in units of the critical 
magnetic field 
\beq
B_{\rm crit} = \left(\frac{m_ec^2}{\alpha \lambda_e^3}\right)^{1/2} \simeq 4.4 \times 10^{13} {\rm G} \ , 
\eeq
$\lambda_e=\hbar/(m_e c)$ being the electron Compton wavelength and $\alpha=e^2/(\hbar c)$ the fine structure constant, 
the baryon density and pressure at neutron drip are approximately given by \cite{chamel2015a}
\beq
\label{eq:ndrip-cat-approx}
n_{\rm drip}(A,Z) \approx \frac{A}{Z} \frac{ \mu_e^{\rm drip}(A,Z)^3}{3\pi^2 (\hbar c)^3}
\biggl[1+\frac{4 C \alpha}{(81\pi^2)^{1/3}} Z^{2/3} \biggr]^{-3}\, .
\eeq
\beq
\label{eq:Pdrip-cat-approx}
P_{\rm drip}(A,Z) \approx \frac{\mu_e^{\rm drip}(A,Z)^4}{12 \pi^2 (\hbar c)^3}\biggl[1+\frac{4C \alpha Z^{2/3}}{(81\pi^2)^{1/3}} \biggr]^{-3} \, .
\eeq
In the presence of a strongly quantizing magnetic field such that 
$n_e<n_{e\text{B}}$ and $T<T_\text{B}$ with 
\begin{equation}\label{eq:neB}
n_{e\text{B}}= \frac{B_\star^{3/2}}{\sqrt{2} \pi^2 \lambda_e^3} \, ,
\end{equation}
\begin{equation}\label{eq:TB}
 T_\text{B}=\frac{m_e c^2}{k_\text{B}} B_\star\, , 
\end{equation}
the neutron-drip density and pressure are approximately given by \cite{chamel2015b}
\beq
\label{eq:ndrip-cat-approx-b}
n_{\rm drip}(A,Z) \approx \frac{A}{Z} \frac{ \mu_e^{\rm drip}(A,Z)}{m_e c^2} \frac{B_\star}{2\pi^2 \lambda_e^3}
\biggl[1-\frac{4}{3} C \alpha Z^{2/3} \left(\frac{B_\star}{2 \pi^2}\right)^{1/3} \left(\frac{m_e c^2}{\mu_e^{\rm drip}}\right)^{2/3} \biggr] \, ,
\eeq
\beq
\label{eq:Pdrip-cat-approx-b}
P_{\rm drip}(A,Z) \approx \frac{B_\star \mu_e^{\rm drip}(A,Z)^2}{4 \pi^2 \lambda_e^3 m_e c^2}\biggl[1 - \frac{1}{3} C \alpha Z^{2/3} \left(\frac{4 B_\star}{\pi^2}\right)^{1/3} \left(\frac{m_e c^2}{\mu_e^{\rm drip}}\right)^{2/3} \biggr] \, .
\eeq
Using Eqs.~(\ref{eq:neB}) and (\ref{eq:ndrip-cat-approx-b}), the condition $n_e<n_{e\rm B}$ can be approximately expressed to first order in $\alpha$  as 
\begin{equation}\label{eq:Bdrip}
B_\star > B_\star^{\rm drip}\equiv \frac{1}{2}\left(\frac{ \mu_e^{\rm drip}(A,Z)}{m_e c^2}\right)^2 
\biggl[1-\frac{8 C \alpha Z^{2/3}}{3(2 \pi^2)^{1/3}} \biggr] \, .
\end{equation}

\subsection{Accreting neutron stars}

In accreting neutron stars, the magnetic field is typically negligibly small ($B\ll B_\star$) and will thus be ignored.  
For an accretion rate $\dot{M}=10^{-9}~{\rm M_\odot}$~yr$^{-1}$ the original outer crust is replaced by accreted matter in $10^4$~yr.
For low-mass binary systems, the accretion stage can last for $10^9$~yr.
At densities above $\sim 10^8$~g~cm$^{-3}$, matter is highly degenerate and relatively cold  ($T\lesssim 5\times 10^8$~K) so that thermonuclear processes are strongly suppressed, since their rates are many orders of magnitude lower than the compression rate due to accretion \cite{haensel2007}. The only relevant 
processes are electron captures and neutron-emission processes, whereby the nucleus $^A_ZX$ is transformed into a nucleus $^{A-\Delta N}_{Z-1}Y$ with 
proton number $Z-1$ and mass number $A-\Delta N$ by capturing an electron with the emission of $\Delta N$ neutrons $n$ and an electron neutrino $\nu_e$:
\beq
\label{eq:e-capture+n-emission}
^A_ZX+ e^- \rightarrow ^{A-\Delta N}_{Z-1}Y+\Delta N n+ \nu_e\, .
\eeq
In this case, the condition for the onset of neutron drip becomes~\cite{chamel2015a}
\begin{eqnarray}
\label{eq:e-capture+n-emission-gibbs-approx}
\mu_e + C e^2 n_e^{1/3}\biggl[Z^{5/3}-(Z-1)^{5/3} + \frac{1}{3} Z^{2/3}\biggr] = \mu_e^{\rm drip-acc} \, ,
\end{eqnarray}
where
\begin{equation}
\label{eq:muebetan}
\mu_e^{\rm drip-acc}(A,Z)\equiv M'(A-\Delta N,Z-1)c^2-M'(A,Z)c^2 +m_n c^2 \Delta N + m_e c^2 \, .
\end{equation}
The neutron-drip density and pressure are approximately given by~\cite{chamel2015a}
\begin{equation}
\label{eq:ndrip-acc}
n_{\rm drip-acc}(A,Z) \approx \frac{A}{Z} \frac{\mu_e^{\rm drip-acc}(A,Z)^3}{3\pi^2 (\hbar c)^3} 
 \biggl[1+\frac{C \alpha}{(3\pi^2)^{1/3}}\left(Z^{5/3}-(Z-1)^{5/3}+\frac{Z^{2/3}}{3}\right)\biggr]^{-3}\, ,
\end{equation}
\begin{equation}
\label{eq:Pdrip-acc}
P_{\rm drip-acc}(A,Z) \approx \frac{\mu_e^{\rm drip-acc}(A,Z)^4}{12 \pi^2 (\hbar c)^3}\biggl[1+\frac{4C \alpha Z^{2/3}}{(81\pi^2)^{1/3}} \biggr]
 \biggl[1+\frac{C \alpha}{(3\pi^2)^{1/3}}\left(Z^{5/3}-(Z-1)^{5/3}+\frac{Z^{2/3}}{3}\right)\biggr]^{-4}\, .
\end{equation}
As discussed in Ref.~\cite{chamel2015a}, the dripping nucleus can be determined as follows. Given the mass number $A$ and the initial atomic number 
$Z_0$ of the ashes of x-ray bursts, the atomic number $Z$ at the neutron-drip point is the highest number of protons lying below $Z_0$ for which the $\Delta N$-neutron 
separation energy defined as
\beq
\label{eq:sn-dn}
S_{\Delta N n}(A,Z-1) \equiv M(A-\Delta N,Z-1)c^2 - M(A,Z-1)c^2 + \Delta N m_n c^2
\eeq
is negative.

\section{Numerical results}
\label{sec:results}

\subsection{Nonaccreting neutron stars}
\label{sec:results-nonaccreting}

We have calculated the properties of neutron-star crusts at the neutron-drip point by minimizing the Gibbs free energy per nucleon~(\ref{eq:gibbs}), 
both in the absence and in the presence of a strong magnetic field. In the latter case, we have set $B_\star=500$, $1000$, $1500$ and $2000$ corresponding 
to magnetic fields in the range $2.2 \times 10^{16}$~G to $8.8 \times 10^{16}$~G. Since nuclear masses at this depth of the outer crust are not experimentally 
known, the predictions for the dripping nucleus are model dependent. The neutron-drip properties are summarized in Table~\ref{tab:drip-cat} in 
unmagnetized neutron stars, and in Tables~\ref{tab:drip-cat-mag-500}-\ref{tab:drip-cat-mag-2000} in strongly magnetized neutron stars (magnetars). 

\begin{figure*}
\begin{center}
\includegraphics[scale=0.45]{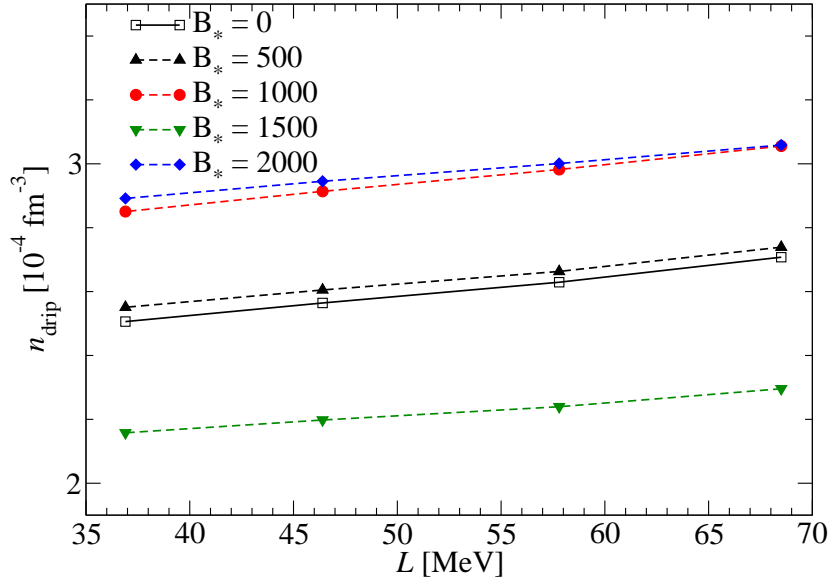}
\end{center}
%\vskip -0.5cm
\caption{(Color online) Neutron-drip density in nonaccreting neutron-star crusts as a function of the slope $L$ of the symmetry energy of infinite homogeneous nuclear matter at saturation
and for different magnetic field
strengths, as obtained using the HFB-22 to HFB-25 Brussels-Montreal nuclear mass models~\cite{goriely2013}.}
\label{fig:cat-ndrip-B}
\end{figure*}

\begin{figure*}
\begin{center}
\includegraphics[scale=0.45]{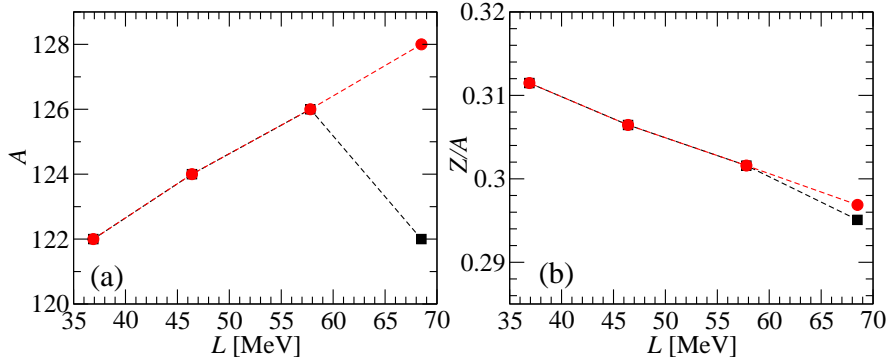}
\end{center}
%\vskip -0.5cm
\caption{(Color online) (a) Mass number $A$ and (b) proton fraction $Z/A$ at the neutron-drip transition for nonaccreting neutron-star crusts as a function of the slope $L$ of the 
symmetry energy of infinite homogeneous nuclear matter at saturation, 
as obtained using the HFB-22 to HFB-25 Brussels-Montreal nuclear mass models~\cite{goriely2013}. Squares (circles) correspond to $B_\star=0, 500$ and $1500$ 
($B_\star =1000$ and $2000$).}
\label{fig:cat-AZ-B}
\end{figure*}

Figure~\ref{fig:cat-ndrip-B} shows that for any given value of the magnetic field strength, the neutron-drip density increases almost linearly with the slope 
of the symmetry energy $L$ (or equivalently with $J$ since the two coefficients are strongly correlated, as previously discussed in Sec.~\ref{sec:hfb-models}). 
On the other hand, the behavior of the neutron-drip density with respect to the magnetic field strength exhibits typical quantum oscillation whereas the 
neutron-drip pressure increases monotonically, as recently discussed in Ref.~\cite{chamel2015b}. The errors of the analytical formulas~(\ref{eq:ndrip-cat-approx})-(\ref{eq:Pdrip-cat-approx}) amount to $0.1\%$ at most, as compared to the numerical solution of Eq.~(\ref{eq:n-drip-mue}). The proton fraction $Z/A$ 
at the neutron-drip point is also found to be strongly correlated with the symmetry energy. As shown in the right panel of Fig.~\ref{fig:cat-AZ-B}, $Z/A$ decreases 
almost linearly with increasing $L$ (or $J$). Similar behaviors of $Z/A$ and $n_{\rm drip}$ with $L$ have been recently obtained in Ref.~\cite{bao2014}, and 
can be inferred from the discussions in Refs.~\cite{roca2008,provid2014,grill2014}. Nevertheless, in all cases they considered the limiting case $B_\star=0$. 
In Ref.~\cite{bao2014}, the authors studied the role of the symmetry energy on the properties of neutron-star crusts around 
the neutron-drip threshold using two sets of relativistic mean field (RMF) models based on the TM1 and IUFSU parametrizations respectively. They generated 
series of models so as to achieve different values of $L$ keeping the symmetry energy at $n=0.11$~fm$^{-3}$ fixed. In our case, the fixed value of the symmetry 
energy at $n\approx 0.11$~fm$^{-3}$ results from the mass fit without any further constraint. Although the variations of $Z/A$ and $n_{\rm drip}$ they found 
are nonlinear over this range of values of $L$, the variations become almost linear on the narrower range we consider (from about 37~MeV to about 69~MeV). 
Although it has been found that a soft symmetry energy favors neutron drip in isolated nuclei~\cite{todd2003}, this result does not necessarily imply the observed 
correlation between $n_{\rm drip}$ and $L$. Indeed, as recently discussed in Ref.~\cite{chamel2015a}, the dripping nucleus in the crust is actually stable 
against neutron emission, but unstable against electron captures followed by neutron emission. Actually, as will be discussed in Sec.~\ref{sec:results-accreting}, 
accreting neutron star crusts exhibit different correlations between $n_{\rm drip}$ and $L$.

The role of the symmetry energy on the properties of the crust at the neutron-drip transition can be understood as follows. Neglecting electron-ion interactions, 
the neutron-drip condition~(\ref{eq:n-drip-mue}) reduces to
\beq
\label{eq:mue-approx}
\mu_e\approx \mu_e^{\rm drip} = m_e c^2 + \frac{A}{Z}\left(m_n c^2- \frac{M^\prime(A,Z)c^2}{A}\right)\, .
\eeq
For the sake of simplicity, let us consider a two-parameter mass formula: 
\beq
\label{eq:2par-ldm}
M^\prime(A,Z)c^2=A \left[a_{\rm eff}+J_{\rm eff} \left(1-2 \frac{Z}{A} \right)^2+m_u c^2 \right] + Z m_e c^2\, ,
\eeq
where $m_u$ is the unified mass unit, $a_{\rm eff}<0$ is the contribution from charge-symmetric matter, while the deviations introduced by 
the charge asymmetry are embedded in the coefficient $J_{\rm eff}>0$. Note that due to nuclear surface effects the values of these coefficients 
do not need to be the same as their corresponding values in infinite homogeneous nuclear matter at saturation. In particular, as already discussed 
in Sec.~\ref{sec:hfb-models}, the ``effective'' symmetry energy coefficient $J_{\rm eff}$ is expected to be smaller than $J$, and to decrease 
with increasing $J$ or $L$. Minimizing $g$ using Eq.~(\ref{eq:mue-approx}) and the mass formula~(\ref{eq:2par-ldm}), the equilibrium proton fraction 
at the neutron-drip transition is approximately given by~\cite{chamel2012} 
\begin{equation}
\label{eq:zovera-cat}
\frac{Z}{A}\approx \frac{1}{2}\sqrt{1+\frac{a_{\rm eff}}{J_{\rm eff}}}\, .
\end{equation}
This shows that $Z/A$ decreases with increasing $L$ (decreasing $J_{\rm eff}$). Note that in this analysis we have not made any assumption 
regarding the magnetic field. In other words, the correlation between $Z/A$ and $L$ (or $J$) is thus expected to be independent of the magnetic 
field strength (at least at the level of accuracy of the simple mass formula considered here), in agreement with the results plotted in the right panel of 
Fig.~\ref{fig:cat-AZ-B}. It follows from Eq.~(\ref{eq:2par-ldm}) that decreasing $J_{\rm eff}$ increases $M^\prime(A,Z)$ (the energy cost associated 
with charge asymmetry is reduced, therefore nuclei are more bound). Using Eq.~(\ref{eq:mue-approx}), we find that $\mu_e^{\rm drip}$ increases with 
$L$. Since $n_{\rm drip}$ and $P_{\rm drip}$ increase with $\mu_e^{\rm drip}$, as shown in Eqs.~(\ref{eq:ndrip-cat-approx}) and (\ref{eq:Pdrip-cat-approx})
in the absence of magnetic field, and in Eqs.~(\ref{eq:ndrip-cat-approx-b}) and (\ref{eq:Pdrip-cat-approx-b}) in the presence of a strongly quantizing 
magnetic field, we can thus conclude that the neutron-drip transition is shifted to higher density and pressure with increasing the symmetry energy, as 
shown in Fig.~\ref{fig:cat-ndrip-B}.

\begin{figure*}
\begin{center}
\includegraphics[scale=0.45]{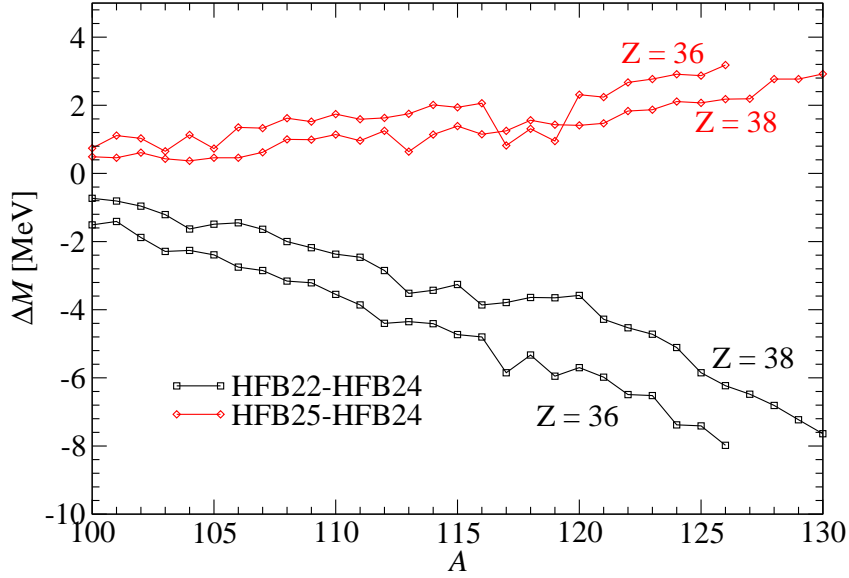}
\end{center}
%\vskip -0.5cm
\caption{(Color online) Difference in mass predictions for two pairs of Brussels-Montreal nuclear mass models along the two isotopic chains $Z=36$ and $Z=38$ relevant at neutron drip.}
\label{fig:mass_diff}
\end{figure*}

The equilibrium nucleus at the neutron-drip transition is less sensitive to the symmetry energy, as previously noticed in Ref.~\cite{bao2014} in the absence 
of magnetic fields (see their Fig.~9). This can be understood as follows. The equilibrium with respect to weak interaction processes requires 
\beq
\label{eq:betaeq}
\mu_p+\mu_e=\mu_n\, ,
\eeq
where $\mu_p$ ($\mu_n$) is the proton (neutron) chemical potential. Substituting the neutron-drip value of the neutron chemical potential $\mu_n=m_n c^2$ in
Eq.~(\ref{eq:betaeq}) and using Eq.~(\ref{eq:mue-approx}), we obtain
\begin{equation}
\label{eq:mup}
\mu_p-m_p c^2 = Q_{n,\beta}+\frac{A}{Z}\left(\frac{M^\prime(A,Z)c^2}{A}-m_n c^2 \right)\, ,
\end{equation}
where $m_p$ is the proton mass and $Q_{n,\beta} = 0.782$~MeV is the $\beta$-decay energy of the neutron. The quantity on the left-hand side of Eq.~(\ref{eq:mup}) 
is approximately equal to the opposite of the one-proton separation energy. This shows that the equilibrium nucleus is uniquely determined by nuclear masses only, 
and is sensitive to the details of the nuclear structure. As a consequence, the predicted nucleus depends on the nuclear mass model employed (see, e.g., 
Refs.~\cite{roca2008,pearson2011,wolf2013,kreim2013,chamel2015c}). 
To better illustrate this point, we have plotted in Fig.~\ref{fig:mass_diff} the differences in 
the mass predictions between HFB-22, HFB-25, and HFB-24 mass models for two isotopic chains, corresponding to the proton number at the neutron-drip 
point (see also Table~\ref{tab:drip-cat}). As shown in Fig.~\ref{fig:mass_diff}, the HFB-22 model deviates more significantly from the ``reference'' 
model HFB-24 than HFB-25, thus explaining the quantitative differences in the dripping nucleus. The variations of $Z$ and $A$ with $L$ we find 
appear to be more irregular than those shown in Fig.~9 of Ref.~\cite{bao2014}. This stems from the fact that in Ref.~\cite{bao2014} nuclear masses were 
calculated using the semi-classical Thomas-Fermi approximation, which does not take into account pairing and shell effects contrary to the fully quantum mechanical 
mass models~\cite{goriely2013} employed here.

\begin{figure*}
\begin{center}
\includegraphics[scale=0.45]{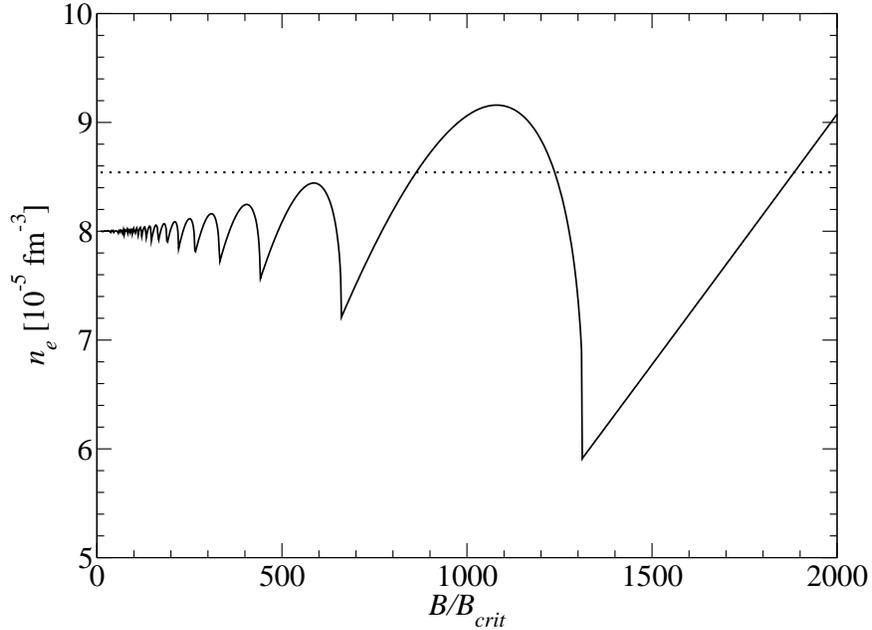}
\end{center}
%\vskip -0.5cm
\caption{Electron density $n_e$ (solid line) at the neutron-drip transition in nonaccreting neutron-star crusts as a 
function of the magnetic field strength, using the Brussels-Montreal nuclear mass model HFB-22 and assuming that the neutron-drip nucleus 
is $^{122}$Kr, as in the absence of magnetic field. The horizontal dotted line represents $n_e^0$, as given by Eq.~(\ref{eq:ne0}). 
See text for details. }
\label{fig:drip-nuc-hfb22}
\end{figure*}

The presence of a strong magnetic field can change the composition at the neutron-drip point, as shown in the left panel of Fig.~\ref{fig:cat-AZ-B} 
(see also Tables~\ref{tab:drip-cat-mag-500}-\ref{tab:drip-cat-mag-2000}). 
However, such behavior is only observed for the nuclear mass model HFB-22. In particular, the equilibrium nucleus is $^{122}$Kr for $B_\star=0, 500$ and $1500$, 
while for $B_\star=1000$ and $2000$ it is $^{128}$Sr. These results can be understood as follows. As discussed in Sec.~\ref{sec:neutron-drip}, the equilibrium 
nucleus at the neutron-drip pressure $P_{\rm drip}$ must be such as to minimize the Gibbs free energy per nucleon, therefore we must have 
\begin{equation}\label{eq:stability}
g(A,Z,P_{\rm drip}) < g(A^\prime,Z^\prime,P_{\rm drip})\, ,
\end{equation}
for any values of $A^\prime\neq A$ and $Z^\prime\neq Z$. This condition can be approximately expressed as~\cite{chamel2015b}
\begin{equation}\label{eq:dripping-nucleus}
\frac{M^\prime(A^\prime,Z^\prime)c^2}{Z^\prime} - \frac{M^\prime(A,Z)c^2}{Z} > \left(\frac{A^\prime}{Z^\prime}-\frac{A}{Z}\right) m_n c^2 + C e^2 n_e^{1/3}\left( Z^{2/3}-Z^{\prime\, 2/3}\right)\, ,
\end{equation}
where the electron density $n_e$ has to be determined from Eq.~(\ref{eq:n-drip-mue}). Equation~(\ref{eq:dripping-nucleus}) can be equivalently written 
as $n_e < n_e^0$, where 
\begin{equation}\label{eq:ne0}
n_e^0\equiv \biggl[\frac{M^\prime(A^\prime,Z^\prime)c^2}{Z^\prime} - \frac{M^\prime(A,Z)c^2}{Z} - \left(\frac{A^\prime}{Z^\prime}-\frac{A}{Z}\right) m_n c^2\biggr]^3
\biggl[C e^2 \left( Z^{2/3}-Z^{\prime\, 2/3}\right)\biggr]^{-3}\, .
\end{equation}
The HFB-22 nuclear mass model predicts very similar values for the  
threshold electron Fermi energy $\mu_e^{\rm drip}$ for nuclei $^{128}$Sr and $^{122}$Kr: $24.970$ and $25.006$~MeV respectively. Substituting the theoretical 
values of the masses of $^{128}$Sr and $^{122}$Kr in Eq.~(\ref{eq:ne0}) with $Z=36$, $A=122$, $Z^\prime=38$, $A^\prime=128$, we obtain $n_e^0\approx 8.54\times 10^{-5}$~fm$^{-3}$. 
Due to Landau quantization of electron motion, $n_e$ varies non-monotonically with $B_\star$. As a consequence, the lattice term in Eq.~(\ref{eq:dripping-nucleus}) 
can thus become comparable to the other terms depending on $B_\star$ to the effect that the condition (\ref{eq:stability}) may be violated (i.e. $n_e\geq n_e^0$), as 
shown in Fig.~\ref{fig:drip-nuc-hfb22}. 
Transitions between $^{128}$Sr and $^{122}$Kr are found to occur at magnetic field strengths $B_\star\approx 861, 1239$ and $1883$. As shown in the right panel of 
Fig.~\ref{fig:cat-AZ-B}, the proton fraction $Z/A$ is barely affected by these changes of composition. In other words, the correlation between $Z/A$ and the symmetry 
energy is almost independent of the magnetic field strength, as previously discussed.

\begin{table}
\centering
\caption{Neutron-drip transition in the crust of nonaccreting and unmagnetized neutron stars, as predicted by the HFB-22 to HFB-25 Brussels-Montreal nuclear mass models: mass and atomic numbers of the dripping nucleus, baryon density, and corresponding pressure. }\smallskip
\label{tab:drip-cat}
\begin{tabular}{ccccc}
\hline % \noalign {\smallskip}
 & $A$ & $Z$ & $n_{\rm drip}$ ($10^{-4}$ fm$^{-3}$) & $P_{\rm drip}$ ($10^{-4}$ MeV~fm$^{-3}$)\\
\hline \noalign {\smallskip}
HFB-22 & 122 & 36 & 2.71 & 4.99  \\
HFB-23 & 126 & 38 & 2.63 & 4.93  \\
HFB-24 & 124 & 38 & 2.56 & 4.87  \\
HFB-25 & 122 & 38 & 2.51 & 4.83 \\
\hline
\end{tabular}
\end{table}

\begin{table}
\centering
\caption{Neutron-drip transition in the crust of nonaccreting magnetized neutron stars with $B_\star=500$, as predicted by the HFB-22 to HFB-25 Brussels-Montreal nuclear mass models: mass and atomic numbers of the dripping nucleus, baryon density, and corresponding pressure.}\smallskip
\label{tab:drip-cat-mag-500}
\begin{tabular}{ccccc}
\hline % \noalign {\smallskip}
 & $A$ & $Z$ & $n_{\rm drip}$ ($10^{-4}$ fm$^{-3}$) & $P_{\rm drip}$ ($10^{-4}$ MeV~fm$^{-3}$)\\
\hline \noalign {\smallskip}
HFB-22 & 122 & 36 & 2.74 & 5.52 \\
HFB-23 & 126 & 38 & 2.66 & 5.45 \\
HFB-24 & 124 & 38 & 2.61 & 5.39 \\
HFB-25 & 122 & 38 & 2.55 & 5.35 \\
\hline
\end{tabular}
\end{table}

\begin{table}
\centering
\caption{Same as in Table~\ref{tab:drip-cat-mag-500} but for $B_\star=1000$.}\smallskip
\label{tab:drip-cat-mag-1000}
\begin{tabular}{ccccc}
\hline % \noalign {\smallskip}
 & $A$ & $Z$ & $n_{\rm drip}$ ($10^{-4}$ fm$^{-3}$) & $P_{\rm drip}$ ($10^{-4}$ MeV~fm$^{-3}$)\\
\hline \noalign {\smallskip}
HFB-22 & 128 & 38 & 3.06 & 6.70 \\
HFB-23 & 126 & 38 & 2.98 & 6.63 \\
HFB-24 & 124 & 38 & 2.91 & 6.56 \\
HFB-25 & 122 & 38 & 2.85 & 6.52 \\
\hline
\end{tabular}
\end{table}

\begin{table}
\centering
\caption{Same as in Table~\ref{tab:drip-cat-mag-500} but for $B_\star=1500$.}\smallskip
\label{tab:drip-cat-mag-1500}
\begin{tabular}{ccccc}
\hline % \noalign {\smallskip}
 & $A$ & $Z$ & $n_{\rm drip}$ ($10^{-4}$ fm$^{-3}$) & $P_{\rm drip}$ ($10^{-4}$ MeV~fm$^{-3}$)\\
\hline \noalign {\smallskip}
HFB-22 & 122 & 36 & 2.30 & 8.66 \\
HFB-23 & 126 & 38 & 2.24 & 8.60 \\
HFB-24 & 124 & 38 & 2.20 & 8.56 \\
HFB-25 & 122 & 38 & 2.16 & 8.52 \\
\hline
\end{tabular}
\end{table}

\begin{table}
\centering
\caption{Same as in Table~\ref{tab:drip-cat-mag-500} but for $B_\star=2000$.}\smallskip
\label{tab:drip-cat-mag-2000}
\begin{tabular}{ccccc}
\hline % \noalign {\smallskip}
 & $A$ & $Z$ & $n_{\rm drip}$ ($10^{-4}$ fm$^{-3}$) & $P_{\rm drip}$ ($10^{-3}$ MeV~fm$^{-3}$)\\
\hline \noalign {\smallskip}
HFB-22 & 128 & 38 & 3.06 & 11.6 \\
HFB-23 & 126 & 38 & 3.00 & 11.6 \\
HFB-24 & 124 & 38 & 2.95 & 11.5 \\
HFB-25 & 122 & 38 & 2.89 & 11.4 \\
\hline
\end{tabular}
\end{table}

\subsection{Accreting neutron stars}
\label{sec:results-accreting}

For accreting neutron-star crusts, we have studied the neutron-drip transition as explained in Sec.~\ref{sec:neutron-drip} (see also Ref.~\cite{chamel2015a}). 
We have considered different initial compositions: the ashes produced by an $rp$-process during an x-ray burst \cite{schatz2001}, and the ashes produced by 
steady state hydrogen and helium burning \cite{schatz2003} as expected to occur during superbursts \cite{gupta2007}. After determining the dripping nucleus, we 
have calculated the neutron-drip density and pressure by solving numerically Eq.~(\ref{eq:e-capture+n-emission-gibbs-approx}) considering all possible 
neutron-emission processes. Results are summarized in Tables~\ref{tab:drip-acc-22}-\ref{tab:drip-acc-25} for different nuclear mass models. 
For comparison with previous works~\cite{hz1990, hz2003}, we have also considered ashes of x-ray bursts consisting of pure $^{56}$Fe. Results are indicated in 
Table~\ref{tab:drip-acc-a56}. 

As already pointed out in Ref.~\cite{chamel2015a}, for a given nuclear mass model the neutron-drip transition in accreting neutron stars can occur at 
either lower or at higher densities and pressures than in nonaccreting neutron stars. Depending on the mass model adopted, the neutron-drip density thus 
ranges from $1.60 \times 10^{-4}$~fm$^{-3}$ to $3.90 \times 10^{-4}$~fm$^{-3}$, 
and the corresponding pressure from $2.77 \times 10^{-4}$~MeV~fm$^{-3}$ to $7.77 \times 10^{-4}$~MeV~fm$^{-3}$. The numerical results 
obtained solving Eq.~(\ref{eq:e-capture+n-emission-gibbs-approx}) are reproduced by the analytical formulas~(\ref{eq:ndrip-acc}) and 
(\ref{eq:Pdrip-acc}) with an error of $0.1\%$ at most. 

\begin{figure*}
\begin{center}
\includegraphics[scale=0.45]{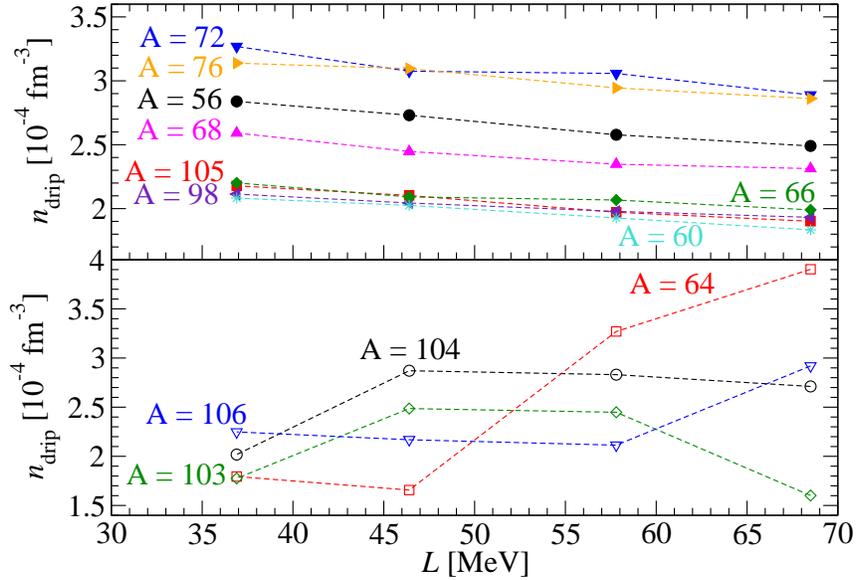}
\end{center}
%\vskip -0.5cm
\caption{(Color online) Neutron-drip density as a function of the slope $L$ of the symmetry energy of infinite homogeneous nuclear matter at saturation, as obtained using the HFB-22 to HFB-25 Brussels-Montreal nuclear mass 
models, for accreting neutron-star crusts with different initial composition of ashes (see text for details).}
\label{fig:acc-ndrip}
\end{figure*}

The change of the neutron-drip density with the slope $L$ of the symmetry energy is found to be very different from that obtained in 
nonaccreting neutron-star crusts. As shown in Fig.~\ref{fig:acc-ndrip}, for some ashes like $^{104}$Cd, no obvious correlation is observed while for 
other ashes like $^{66}$Ni, $n_{\rm drip-acc}$ appears to be \emph{anticorrelated} with $L$: $n_{\rm drip-acc}$ decreases with increasing $L$. 
This behavior can be understood as follows. Ignoring electron-ion interactions, the neutron-drip condition~(\ref{eq:e-capture+n-emission-gibbs-approx}) 
reduces to
\beq
\mu_e\approx \mu_e^{\rm drip-acc} = M^\prime(A-\Delta N,Z-1)c^2-M^\prime(A,Z)c^2 +m_n c^2 \Delta N + m_e c^2\, , 
\eeq
which can be more conveniently written as 
\beq\label{eq:muedrip-acc-s}
\mu_e^{\rm drip-acc} = S_{\Delta N n}(A,Z-1) + \mu_e^\beta(A,Z)\, ,
\eeq
using Eq.~(\ref{eq:sn-dn}) and introducing the threshold electron Fermi energy for the onset of electron captures (see, e.g. Ref.~\cite{chamel2015d} for a recent 
discussion)
\beq
\mu_e^\beta(A,Z)=M^\prime(A,Z-1)c^2 -M^\prime(A,Z)c^2 + m_e c^2\, .
\eeq
The mass difference $\Delta M^\prime=M'(A,Z-1) -M'(A,Z)$, which represents the change of mass associated with the substitution of a proton by a neutron, 
is expected to be mainly determined by symmetry energy effects. On the other hand, the $\Delta N$-neutron separation energy $S_{\Delta N n}(A,Z-1)$ 
is likely to be more dependent on the details of the nuclear structure than on the symmetry energy. As shown in Tables~\ref{tab:drip-acc-22}-\ref{tab:drip-acc-a56}, 
$|S_{\Delta N n}(A,Z-1)| \ll \Delta M^\prime c^2$ therefore $\mu_e^{\rm drip-acc}\approx \mu_e^\beta$. On the other hand, as discussed in Sec.~\ref{sec:neutron-drip}, 
the composition of accreting neutron-star crusts at the neutron-drip transition is directly determined by the condition $S_{\Delta N n}(A,Z-1)<0$. Provided 
the dependence on the symmetry energy of $S_{\Delta N n}(A,Z-1)$ is weak enough, the dripping nucleus will thus be independent of $L$. In this case, the 
variations of $S_{\Delta N n}(A,Z-1)$ with $L$ are typically much smaller than the variations of $\Delta M^\prime c^2$, as shown in Figs.~\ref{fig:deltamprime} and 
\ref{fig:Sdeltan}. Using the simple mass formula~(\ref{eq:2par-ldm}), we find
\beq
\mu_e^\beta(A,Z)=\Delta M^\prime c^2 +m_e c^2 \approx 4J_{\rm eff}\left(1+\frac{1-2Z}{A}\right)\, .
\eeq
This means that with increasing $J$ or $L$ (decreasing $J_{\rm eff}$), $\Delta M^\prime$ and $\mu_e^\beta(A,Z)$ both \emph{decrease}. It thus follows from 
Eqs.~(\ref{eq:ndrip-acc}), (\ref{eq:Pdrip-acc}), and Eqs.~(\ref{eq:muedrip-acc-s}), that $\mu_e^{\rm drip-acc}$, $n_{\rm drip-acc}$ and $P_{\rm drip-acc}$ 
also decrease with $L$, as shown in the upper panel of Fig.~\ref{fig:acc-ndrip}. 
In the peculiar case of $A=105$, the rather low value of $\Delta M^\prime c^2$ predicted by HFB-25 is compensated by a comparatively high value $S_{\Delta N n}(A,Z-1)$, 
as can be seen in Figs.~\ref{fig:deltamprime} and \ref{fig:Sdeltan}. As a result, $n_{\rm drip-acc}$ is still found to decrease with increasing $L$ despite the nonmonotonic 
variation of $\Delta M^\prime$, as shown in Fig.~\ref{fig:acc-ndrip}. 
For some ashes, the variations of $S_{\Delta N n}(A,Z-1)$ are comparable to those of $\Delta M^\prime c^2$, and large enough to even change the composition. This  
leads to nonmonotonic variations of the neutron-drip density and pressure, as illustrated in the lower panel of Fig.~\ref{fig:acc-ndrip}. In these cases, 
effects other than the symmetry energy play a role. The anticorrelation between $n_{\rm drip-acc}$ (or $P_{\rm drip-acc}$) and $L$ thus relies to 
a large extent on the importance of nuclear structure effects far from the stability valley. 

\begin{table}
\centering
\caption{Neutron-drip transition in the crust of accreting neutron stars, as predicted by the HFB-22 Brussels-Montreal nuclear mass model: mass and atomic numbers of the dripping nucleus, number of emitted neutrons, baryon density $n_{\rm drip-acc}$  ($10^{-4}$ fm$^{-3}$),  and corresponding pressure $P_{\rm drip-acc}$ ($10^{-4}$ MeV~fm$^{-3}$), $S_{\Delta N n}(A,Z-1)$ (MeV), $\Delta M^\prime$ (MeV$/c^2$), and $\mu_e^{\rm drip-acc}$ (MeV).
The mass numbers $A$ are listed from top to bottom considering that the ashes are produced by ordinary
x-ray bursts (upper panel) or superbursts (lower panel). See text for details.}\smallskip
\label{tab:drip-acc-22}
\begin{tabular}{cccccccc}
\hline %\noalign {\smallskip}
  $A$   & $ Z$  & $\Delta N$&  $n_{\rm drip-acc}$ ($10^{-4}$ fm$^{-3}$)  & $P_{\rm drip-acc}$ ($10^{-4}$ MeV~fm$^{-3}$) & $S_{\Delta N n}$ (MeV) & $\Delta M^\prime$ (MeV$/c^2$) & $\mu_e^{\rm drip-acc}$ (MeV) \\
 \hline %\noalign {\smallskip}
 104  &   32 &  1 &   2.71 &  5.31 & -0.79 & 25.14 & 24.86 \\
 105  &   33 &  1 &   1.90 &  3.40 & -1.00 & 22.70 & 22.21 \\
  68  &   22 &  1 &   2.31 &   4.64 & -0.28 & 24.14 & 24.37 \\
  64  &   18 &  5 &   3.90 &   7.77 & -1.85 & 29.23 & 27.89 \\
  72  &   22 &  1 &   2.89 &   5.78 & -0.31 & 25.55 & 25.75 \\
  76  &   24 &  1 &   2.86 &   5.95 & -0.17 & 25.52 & 25.86 \\
  98  &   32 &  1 &   1.93 &   3.66 & -0.20 & 22.34 & 22.65 \\
 103  &   33 &  1 &   1.60 &  2.77 & -0.02 & 20.62 & 21.11 \\
 106  &   32 &  1 &   2.92 &  5.71 & -0.69 & 25.50 & 25.32 \\
  \hline
  66  &   22 &  1 &   1.99 &   3.95 & -0.19 & 23.09 & 23.41 \\
  64  &   18 &  5 &   3.90 &   7.77 & -1.85 & 29.23 & 27.89 \\
  60  &   20 &  1 &   1.83 &   3.55 & -1.67 & 24.02 & 22.86 \\
 \end{tabular}
 \end{table}

\begin{table}
\centering
\caption{Same as in Table~\ref{tab:drip-acc-22} but for the HFB-23 Brussels-Montreal nuclear mass model.}\smallskip
\label{tab:drip-acc-23}
\begin{tabular}{cccccccc}
\hline %\noalign {\smallskip}
   $A$   &  $Z$  & $\Delta N$&  $n_{\rm drip-acc}$ ($10^{-4}$ fm$^{-3}$)  & $P_{\rm drip-acc}$ ($10^{-4}$ MeV~fm$^{-3}$) & $S_{\Delta N n}$ (MeV) & $\Delta M^\prime$ (MeV$/c^2$) & $\mu_e^{\rm drip-acc}$ (MeV) \\
 \hline %\noalign {\smallskip}
 104  &   32 &  1 &   2.83 &  5.62 & -1.02 & 25.73 & 25.22 \\
 105  &   33 &  1 &   1.97 &  3.57 & -1.33 & 23.30 & 22.48 \\
  68  &   22 &  1 &   2.35 &   4.73 & -0.56 & 24.54 & 24.49 \\
  64  &   20 &  1 &   3.27 &   7.04 & -0.13 & 26.75 & 27.13 \\
  72  &   22 &  1 &   3.06 &   6.24 & -0.11 & 25.84 & 26.24\\
  76  &   24 &  1 &   2.94 &   6.18 & -0.38 & 25.98 & 26.11 \\
  98  &   32 &  1 &   1.98 &   3.77 & -0.50 & 22.82 & 22.83 \\
 103  &   31 &  1 &   2.45 &  4.50 & -0.91 & 24.29 & 23.89 \\
 106  &   34 &  1 &   2.11 &  4.02 & -0.009 & 22.63 & 23.13 \\
  \hline
  66  &   22 &  1 &   2.07 &   4.16 & -0.23 & 23.43 & 23.71 \\
  64  &   20 &  1 &   3.27 &   7.04 & -0.13 & 26.75 & 27.13 \\
  60  &   20 &  1 &   1.93 &   3.79 & -1.77 & 24.50 & 23.24 \\
 \end{tabular}
 \end{table}

\begin{table}
\centering
\caption{Same as in Table~\ref{tab:drip-acc-22} but for the HFB-24 Brussels-Montreal nuclear mass model.}\smallskip
\label{tab:drip-acc-24}
\begin{tabular}{cccccccc}
\hline %\noalign {\smallskip}
  $A$   &  $Z$  & $\Delta N$&  $n_{\rm drip-acc}$ ($10^{-4}$ fm$^{-3}$)  & $P_{\rm drip-acc}$ ($10^{-4}$ MeV~fm$^{-3}$) & $S_{\Delta N n}$ (MeV) & $\Delta M^\prime$ (MeV$/c^2$) & $\mu_e^{\rm drip-acc}$ (MeV) \\
 \hline %\noalign {\smallskip}
 104  &   32 &  1 &   2.87 &  5.73 & -1.49 & 26.32 & 25.34 \\
 105  &   33 &  1 &   2.10 &  3.89 & -1.11 & 23.57 & 22.97 \\
  68  &   22 &  1 &   2.45 &   5.00 & -0.75 & 25.07 & 24.83 \\
  64  &   22 &  1 &   1.66 &   3.22 & -0.07 & 21.81 & 22.25 \\
  72  &   22 &  1 &   3.08 &   6.28 & -0.29 & 26.07 & 26.29 \\
  76  &   24 &  1 &   3.10 &   6.61 & -0.46 & 26.50 & 26.55 \\
  98  &   32 &  1 &   2.04 &   3.94 & -0.38 & 22.95 & 23.08 \\
 103  &   31 &  3 &   2.49 &  4.59 & -0.89 & 24.39 & 24.01 \\
 106  &   34 &  1 &   2.17 &  4.16 & -1.10 & 23.92 & 23.33 \\
  \hline
  66  &   22 &  1 &   2.09 &   4.22 & -0.29 & 23.58 & 23.80 \\
  64  &   22 &  1 &   1.66 &   3.22 & -0.07 & 21.81 & 22.25 \\
  60  &   20 &  1 &   2.03 &   4.05 & -1.66 & 24.78 & 23.63 \\
 \end{tabular}
 \end{table}

\begin{table}
\centering
\caption{Same as in Table~\ref{tab:drip-acc-22} but for the HFB-25 Brussels-Montreal nuclear mass model.}\smallskip
\label{tab:drip-acc-25}
\begin{tabular}{cccccccc}
\hline %\noalign {\smallskip}
  $A$   &  $Z$  & $\Delta N$&  $n_{\rm drip-acc}$ ($10^{-4}$ fm$^{-3}$)  & $P_{\rm drip-acc}$ ($10^{-4}$ MeV~fm$^{-3}$) & $S_{\Delta N n}$ (MeV) & $\Delta M^\prime$ (MeV$/c^2$) & $\mu_e^{\rm drip-acc}$ (MeV) \\
 \hline %\noalign {\smallskip}
 104  &   34 &  1 &  2.02 &  3.87 & -0.009 & 22.42 & 22.92 \\
 105  &   33 &  1 &  2.18 &  4.07 & -0.57 & 23.30 & 23.24 \\
  68  &   22 &  1 &   2.59 &  5.40 & -0.75 & 25.55 & 25.31 \\
  64  &   22 &  1 &   1.79 &  3.59 & -0.18 & 22.52 & 22.85 \\
  72  &   22 &  1 &   3.27 &  6.82 & -0.55 & 26.87 & 26.83 \\
  76  &   24 &  1 &   3.14 &  6.73 & -0.58 & 26.74 & 26.67 \\
  98  &   32 &  1 &   2.11 &  4.12 & -0.64 & 23.47 & 23.34 \\
 103  &   33 &  1 &  1.78 &  3.18 & -0.76 & 22.10 & 21.85 \\
 106  &   34 &  1 &  2.25 &  4.36 & -0.05 & 23.15 & 23.61 \\
  \hline
  66  &   22 &  1 &   2.20 &   4.52 & -0.22 & 23.92 & 24.21 \\
  64  &   22 &  1 &   1.79 &   3.59 & -0.18 & 22.52 & 22.85 \\
  60  &   20 &  1 &   2.08 &   4.21 & -1.94 & 25.28 & 23.85 \\
 \end{tabular}
 \end{table}

\begin{table}
\centering
\caption{Neutron-drip transition in the crust of accreting neutron stars, as predicted by different Brussels-Montreal nuclear mass models for $^{56}$Fe ashes: atomic number $Z$ of the dripping nucleus, number of emitted neutrons, density and corresponding pressure, $S_{\Delta N n}(A,Z-1)$, $\Delta M^\prime$ (MeV$/c^2$), and $\mu_e^{\rm drip-acc}$ (MeV). See text for details.}\smallskip
\label{tab:drip-acc-a56}
\begin{tabular}{ccccc}
\hline %\noalign {\smallskip}
                                              & HFB-22 & HFB-23  & HFB-24 &  HFB-25    \\
\hline %\noalign {\smallskip}
$Z$                                           &   18   &  18     &  18    &  18   \\
$\Delta N$                                    &    1   &   1     &   3    &  1     \\
$n_{\rm drip-acc}$ ($10^{-4}$ fm$^{-3}$) & 2.49   &  2.58   &  2.73  &  2.84   \\
$P_{\rm drip-acc}$ ($10^{-4}$ MeV~fm$^{-3}$)  & 5.10   &  5.34   &  5.77  &  6.07   \\
$S_{\Delta N n}$ (MeV) & -1.17 & -1.32 & -3.27 & -1.61 \\
$\Delta M^\prime$ (MeV$/c^2$) & 25.76 & 26.20 & 28.64 & 27.32 \\
$\mu_e^{\rm drip-acc}$ (MeV) & 25.10 & 25.39 & 25.88 & 26.22  \\
\end{tabular}
\end{table}

\begin{figure*}
\begin{center}
\includegraphics[scale=0.45]{drip_acc_22-25_deltam.eps}
\end{center}
%\vskip -0.5cm
\caption{(Color online) Mass difference $\Delta M^\prime$ (in units of MeV$/c^2$) as a function of the slope $L$ of the symmetry energy of infinite homogeneous nuclear matter at saturation, as obtained using the 
HFB-22 to HFB-25 Brussels-Montreal nuclear mass models, for accreting 
neutron-star crusts with different initial composition of ashes (see text for details).}
\label{fig:deltamprime}
\end{figure*}

\begin{figure*}
\begin{center}
\includegraphics[scale=0.45]{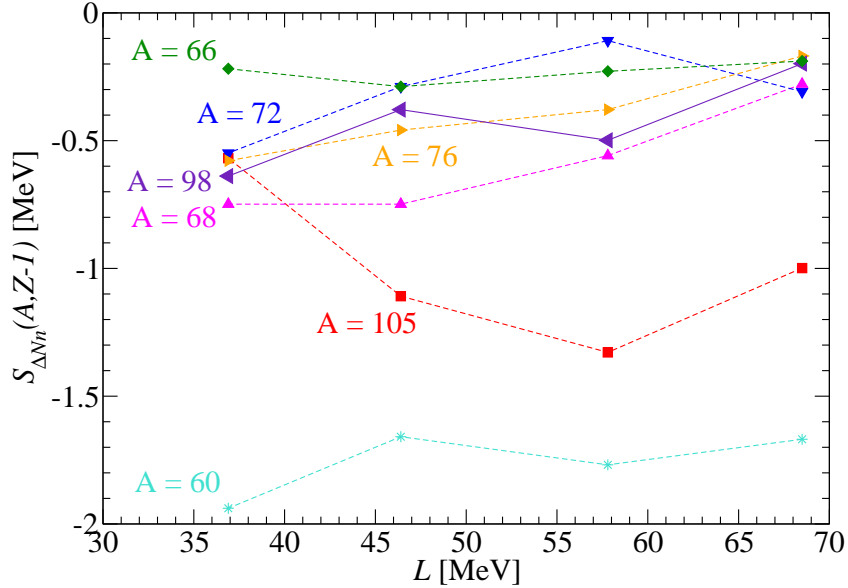}
\end{center}
%\vskip -0.5cm
\caption{(Color online) $\Delta N$-neutron separation energy $S_{\Delta N n}$ (in units of MeV) as a function of the slope $L$ of the symmetry energy of infinite homogeneous nuclear matter at saturation, as obtained using the 
HFB-22 to HFB-25 Brussels-Montreal nuclear mass models, for accreting 
neutron-star crusts with different initial composition of ashes (see text for details).}
\label{fig:Sdeltan}
\end{figure*}

\section{Conclusions}
\label{sec:conclusions}

We have studied the role of the symmetry energy on the neutron-drip transition in both accreting and nonaccreting neutron-star crusts. We have also allowed for  
the presence of a strong magnetic field, as in magnetars. The masses of nuclei encountered in this region of the neutron-star crust are experimentally unknown. 
For this reason, we have employed a recent family of microscopic nuclear mass models, from HFB-22 to HFB-25, developed by the Brussels-Montreal collaboration~\cite{goriely2013}. These models provide equally good fits to the 2353 measured masses of nuclei with $N$ and $Z \geq 8$ from the 2012 Atomic Mass 
Evaluation~\cite{audi2012}, with a root-mean-square deviation of about $0.6$~MeV. 
On the other hand, these models lead to different predictions for the behavior of the symmetry energy in infinite homogeneous
nuclear matter.
In particular, these functionals 
were constrained so as to yield different 
values of the symmetry energy at saturation, from $J=29$~MeV to $J=32$~MeV, the slope of the symmetry energy ranging from $L=37$~MeV to $L=69$~MeV. 

For nonaccreting weakly magnetized neutron stars, the neutron-drip density $n_{\rm drip}$ is found to increase almost linearly with $L$ (or equivalently with $J$) 
while the proton fraction $Z/A$ decreases, in agreement with previous studies~\cite{bao2014} (see also Refs.~\cite{roca2008,provid2014,grill2014}). 
In the presence of a strong magnetic field, the dripping nucleus
hence also $Z/A$ is unchanged, as found in Refs.~\cite{chamel2012,chamel2015b}, for all models but HFB-22. In this case, the dripping nucleus alternates between 
$^{122}$Kr and $^{128}$Sr depending on the magnetic field strength. This peculiar behavior arises from Landau quantization of electron motion and from the fact 
that the threshold electron Fermi energy $\mu_e^{\rm drip}$ are almost equal. Their proton fraction are also very similar so that all in all the linear 
correlation between $Z/A$ and $L$ is hardly affected by the magnetic field. The neutron-drip density $n_{\rm drip}$ exhibits typical quantum oscillations as 
a function of the magnetic field strength, as recently discussed in Ref.~\cite{chamel2015b}. Still, $n_{\rm drip}$ remains linearly correlated with $L$. 
Although a soft symmetry energy favors neutron drip in isolated nuclei~\cite{todd2003}, this result does not necessarily imply the observed correlation 
between $n_{\rm drip}$ and $L$. Indeed, as recently discussed in Ref.~\cite{chamel2015a}, the dripping nucleus in the crust is actually stable 
against neutron emission, but unstable against electron captures followed by neutron emission. In fact, such a correlation is not found in accreting neutron-star crusts. 
Depending on the initial composition of the ashes from x-ray bursts and superbursts, $n_{\rm drip}$ 
\emph{decreases} almost linearly with increasing $L$ while the dripping nucleus remains the same. In other cases, the symmetry energy does not seem 
to play any role. 

We have qualitatively explained these different behaviors using a simple mass formula, and making use of the analytical expressions for the neutron-drip density and 
pressure obtained in Refs.~\cite{chamel2015a,chamel2015b}. In particular, we have shown that the anticorrelation between $n_{\rm drip}$ and $L$ in accreting 
neutron stars depends to a large extent to the relative importance of nuclear structure effects (like shell effects and pairing) and symmetry energy effects on the 
neutron separation energy. More precisely, the anticorrelation is broken whenever the differences between the neutron separation energies predicted by the different 
mass models are large enough to change the dripping nucleus. 

In any case, the composition of the deepest layers of the outer crust of a neutron star is very sensitive to the details of the nuclear structure far from the stability valley. In nonaccreting neutron-star crusts, the neutron-drip transition is mainly governed by the values of the masses of very neutron-rich strontium and krypton isotopes. 
Although the masses of these nuclei have not yet been measured, the composition of nonaccreting neutron-star crusts has been recently constrained by experiment 
to deeper layers~\cite{wolf2013}. The nuclei thought to be present in accreting neutron star crusts span a much larger region of the nuclear chart, depending on 
the ashes from x-ray bursts and superbursts. In these neutron stars, the neutron-drip transition is not directly determined by nuclear masses but rather by some 
combinations of masses, namely the (multiple) neutron separation energies and the isobaric two-point mass differences. 

The onset of neutron emission by nuclei marks the transition to the inner region of the neutron-star crust, where neutron-proton clusters coexist with a neutron liquid. 
In turn, this neutron liquid, which becomes superfluid at low enough temperatures, is expected to play a role in various observed astrophysical phenomena  (see, e.g., 
Ref.~\cite{chamelhaensel2008} for a review) like sudden spin-ups and spin-downs (so-called ``glitches'' and ``antiglitches'', respectively)~\cite{dib2008, gug2014, 
archibald2013, sasmaz2014, duncan2013, kantor2014}, quasiperiodic oscillations detected in the giant flares from soft $\gamma$-ray repeaters~\cite{passamonti2014}, 
cooling of strongly magnetized neutron stars~\cite{aguilera2009}, deep crustal heating (most of the heat being released near the neutron-drip transition~\cite{haensel2008}), 
and the thermal relaxation of quasipersistent soft x-ray transients~\cite{shternin2007,brown2009,page2013}. By shifting the neutron-drip transition to higher or lower densities, the symmetry energy may thus leave its imprint on these astrophysical phenomena.

\begin{acknowledgments}
This work has been mainly supported by Fonds de la Recherche Scientifique - FNRS (Belgium), and by the bilateral project between Fonds de la Recherche Scientifique - FNRS (Belgium), Wallonie-Bruxelles-International (Belgium) and the Bulgarian Academy of Sciences. 
This work has been also partially supported by the Bulgarian National Science Fund under Contract No. DFNI-T02/19t and the Cooperation in Science and Technology (COST Action) MP1304 ``NewCompStar''. The authors would like to thank J.~M. Pearson for very fruitful discussions.
\end{acknowledgments}

\end{document}